 \documentstyle[11pt,aaspp4,flushrt]{article}
\begin{document}
\title{Optical Studies of the Transient Dipping X-ray Sources X1755$-$338 and
 X1658$-$298 in Quiescence}
\author{Stefanie Wachter\altaffilmark{1}}  
\affil{Department of Astronomy, University of Washington, Box 351580,
 Seattle, WA 98195-1580}
\altaffiltext{1}{Visiting Astronomer, Cerro Tololo Interamerican Observatory,
National Optical Astronomy Observatories, operated by AURA, Inc. under 
cooperative agreement with the NSF.}
\authoremail{wachter@astro.washington.edu}
\author{Alan P. Smale\altaffilmark{1}}
\affil{Universities Space Research Association, Code 660.2, NASA/GSFC, 
Greenbelt, MD 20771}
\authoremail{alan@osiris.gsfc.nasa.gov}

\begin{abstract}

We have studied the optical counterparts of X1755$-$338 and X1658$-$298
in their X-ray ``off'' states. The first observations of X1755$-$338 in 
quiescence show that the counterpart, V4134 Sgr, has faded by more 
than 3.5~mag in $V$. 
If the mass donor in the system is an M0V star as implied by the period, 
our upper limits on the brightness of the counterpart suggest that it is more
distant than 4~kpc.

We observed V2134 Oph, the optical counterpart of X1658$-$298,
on several occasions in April/May 1997 and found the
source only $\sim$~1~mag fainter than when
it is X-ray bright. Contemporaneous X-ray data confirm that the source 
remains in the quiescent state during our optical observations. 
Our optical lightcurve, folded on the 7.1~hour
orbital period, does not show any modulation across the binary cycle.
It is possible that the absence of detectable X-ray emission, despite the 
indication 
for an accretion disk and activity in the system, is related to structure 
in the disk that permanently obscures the central X-ray source. 
The optical properties of V2134~Oph are unique among the known X-ray 
transients. 
 
\end{abstract}

\keywords{stars: individual (V4134 Sgr, V2134 Oph) --- 
         stars: variables: other --- X-rays: stars}

\section{Introduction}

About ten systems among the low mass X-ray binaries (LMXBs) exhibit 
irregularly-shaped, recurrent dips
in their X-ray lightcurves. These  
X-ray dippers are believed to be high inclination systems in which 
azimuthal accretion disk
structure extends above the plane of the binary and
periodically blocks the line of sight to the central compact object.
Modeling shows
that for the majority of systems this structure must be at the disk
edge, at the impact point of the accretion stream (White, Nagase \& Parmar 1995
and references therein).
The recurrence time of the dips is assumed to reflect the orbital period. 
The optical/IR emission from dippers also varies on
the orbital period, and originates largely from reprocessing of X-rays
in the outer regions of the system. 

Observations of LMXBs during X-ray quiescence offer the opportunity to
study the optical counterpart ``uncontaminated'' by contributions to the
optical/IR from reprocessed X-radiation in the disk and
secondary. Light from the accretion disk usually dominates in the 
X-ray active state. During quiescence, the mass donor becomes visible which 
allows the determination of its spectral type and radial velocity curve. 
Radial velocity measurements can
provide the mass function and scale of the system and help to
distinguish whether the compact object is a neutron star (NS) or black
hole (BH).

\section{Observations}

CCD $B$, $V$, $R_C$ and $I_C$ band photometry of X1755$-$338 and X1658$-$298
was performed with the CTIO  
1.5~m telescope in April/May 1997 under good seeing conditions 
(1.0--1.2\arcsec). Exposure times ranged from 600-900~s at a
pixel scale of 0.24\arcsec~pix$^{-1}$. 
Overscan and bias corrections were made for each CCD image with the 
task {\it quadproc} at CTIO to deal with the 4 amplifier readout.  
The data were flat-fielded in the standard 
manner using IRAF.
 
Our $V$ band images of the X1755$-$338 and X1658$-$298 fields 
are shown in Figure~\ref{f-fc1} and Figure~\ref{f-fc2}, respectively.  
All comparison stars used in the analysis 
are marked.
Photometry was performed by point spread function fitting with DAOPHOT~II
(Stetson 1993). 
The instrumental magnitudes were transformed to the standard system through
observations of several standard star fields 
(Landolt 1992). 
The resulting magnitudes of X1755$-$338, X1658$-$298 and several local
comparison stars are listed in Table~\ref{t1}. The systematic error (from the 
transformation to the standard system) in these optical	
magnitudes is $\pm 0.10$~mag. For  X1658$-$298, the intrinsic 1$\sigma$ error of 
$\pm 0.02$~mag in 
the relative
photometry was derived from the rms scatter in the lightcurve of comparison
stars of similar brightness. 

\begin{figure}[ht]
\plotfiddle{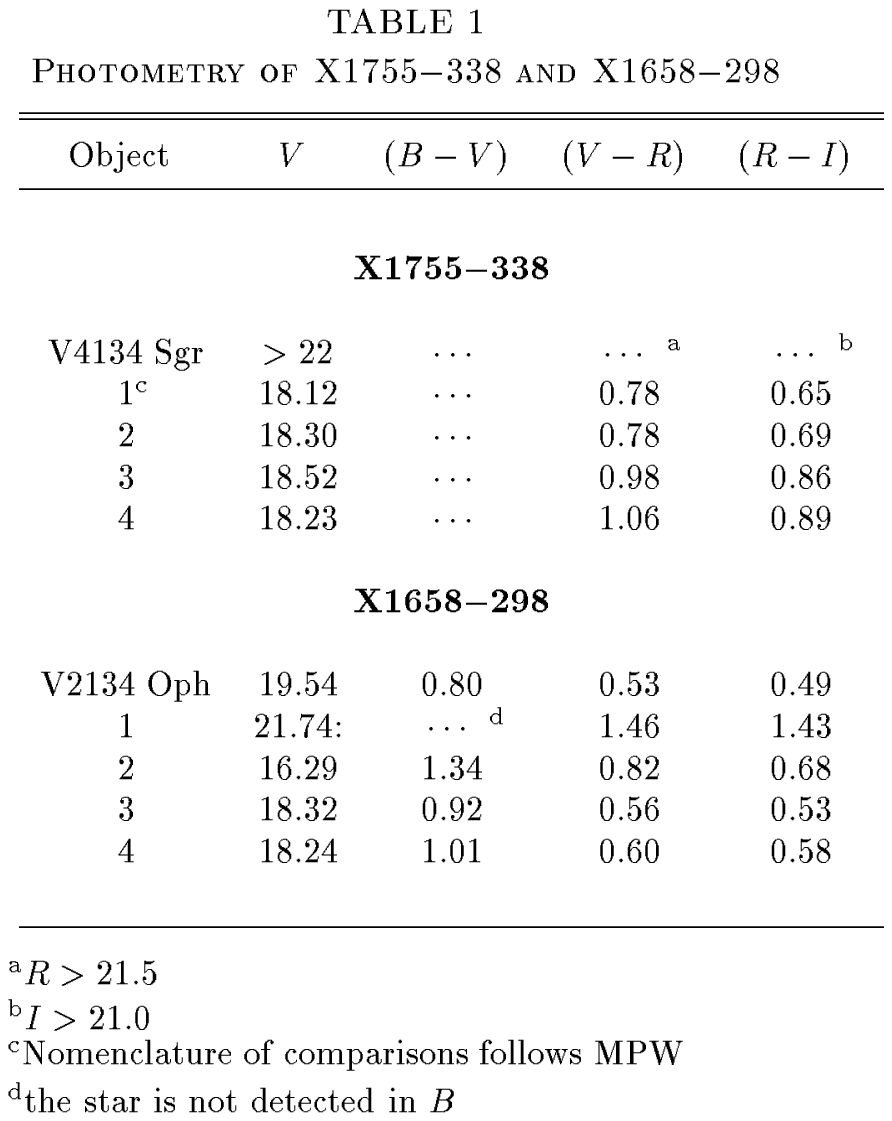}{9.0cm}{0}{80}{80}{-254}{-180}
\end{figure}

\section{Results and Discussion}
 
\subsection{X1755$-$338}

X1755$-$338 was first noted for its very soft X-ray spectrum by Jones (1977). 
Observations with {\it Einstein} identified it as a BH candidate due to its location 
in the ultrasoft region of the X-ray color-color 
diagram (White \& Marshall 1984). 
The {\it Einstein} spectra also indicated a lower than expected 
column density for a source close to the Galactic Bulge.  
{\it EXOSAT} observations by White et al.\ (1984) 
revealed recurrent dips of 30~min duration with a period of 4.4~hours. 
Unlike in other X-ray dippers, the hardness ratio did not change 
between dip and non--dip periods. One explanation offered for this energy independence
of the X-ray dips was a reduction in the metallicity of the absorbing medium
by a factor of 600 
from cosmic abundance values.  
Church \& Balucinska--Church (1993) fitted
the same observations with a two-component model, and were able to reproduce the 
energy-independence of the dips with absorption in material of cosmic abundances.  
They interpret the blackbody component of their model 
as emission from the boundary layer between the accretion disk and the surface 
of a NS, calling into question the BH candidacy of X1755$-$338. 
Pan et al.\ (1995) reported the detection of a hard power-law tail 
in addition to an ultrasoft component in the 
X-ray spectrum of X1755$-$338 and strongly argued in favor of X1755$-$338 being a
BH candidate.  
Simultaneous {\it Ginga} and {\it Rosat} observations confirm the hard tail and 
also show the  
iron 6.7~keV line indicating that the accreted material is {\it not} extremely 
metal-deficient (Seon et al.\ 1995). 

A faint, blue star with a featureless spectrum was suggested as the optical 
counterpart of X1755$-$338 by McClintock et al.\ (1978).  
Mason, Parmar \&
White (1985), hereafter MPW, obtained simultaneous optical and {\it EXOSAT} observations 
of X1755$-$338. Their detection of sinusoidal modulation in the $V$ band with an 
amplitude of
0.4~mag and a period of 4.46~hours confirmed the identification of the counterpart
 (V4134 Sgr).  
The optical minimum occurs 0.15 cycles later than  
the center of the X-ray dip, consistent with the idea that the X-rays 
are being absorbed in material located at the impact point of the accretion stream 
with the accretion disk. 

\begin{figure}[ht]
\epsscale{0.8}
\plotone{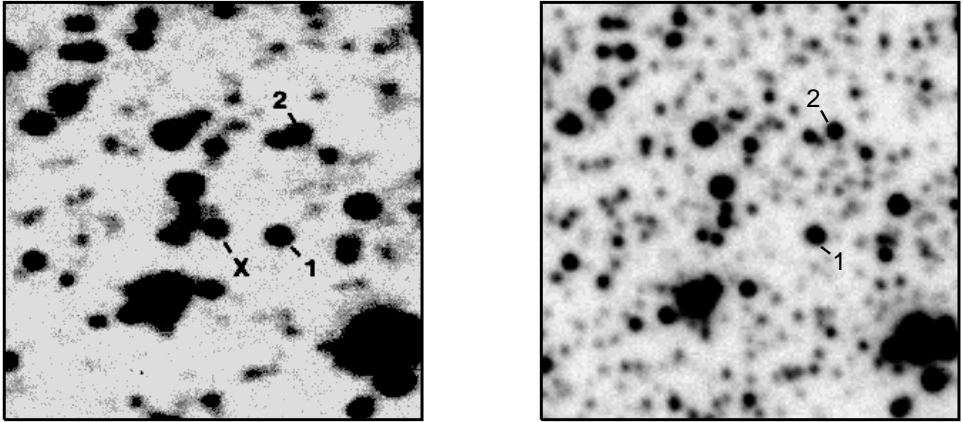}
\caption[]{{\it Left:}  $V$ band exposure of V4134~Sgr 
(X1755$-$338), marked X, during outburst in 1984
(MPW). North is up and east is to the left.
{\it Right:} Our 900~s $V$ band exposure taken on 1997 April 28 UT with the CTIO 1.5~m 
telescope of the same field. Notice that the counterpart of X1755$-$338 has vanished.
\label{f-fc1}}
\end{figure}

In January
1996, X1755$-$338 was 
observed by the Rossi X-ray Timing Explorer (RXTE) in an X-ray off--state 
(a factor of 100 fainter than usual)
for the first time
(Roberts et al.\ 1996).
We obtained optical observations in order to determine the nature 
of the mass donating star. 
The RXTE All Sky Monitor (ASM) lightcurve (available on the Web) 
confirms that the source remained in the off--state 
during our observations.
Figure~\ref{f-fc1} displays our $V$ band observation of the X1755$-$338 field
from 1997 April 28 UT. For comparison, a section of the finding chart 
which shows the source in its bright state 
is also reproduced (with the kind permission of K. Mason).
The counterpart has faded from its on--state brightness of 
$V=18.5$ (MPW) to undetectable magnitudes. 
We obtained upper limits on the counterpart brightness of
$V>22$, $R>21.5$ and $I>21$.
Our $V$ magnitudes for the comparisons 1 through 4 (see Table~\ref{t1})
agree with the measurements
of MPW within the 
uncertainties.  The distance to X1755$-$338 is not well determined.
Estimates in the literature range from 1 to 9~kpc and values for the 
visual extinction lie between $1< A_V < 2$ (MPW). 

In order to derive a distance
from our magnitude limits, we must consider the origin of the optical quiescent
emission in X1755$-$338.  
In quiescence, 
the main light source in the system is most likely the mass donating 
star. There is evidence for a disk even in quiescence in soft X-ray transients, but it
generally does not contribute much of the light in the system
(McClintock \& Remillard 1990; Marsh, Robinson \& Wood 1994).
X1755$-$338 has been X-ray bright for over 20~years, which is very different from 
typical transient behavior (relatively short X-ray outbursts separated by long periods of 
quiescence). For the remainder of the discussion we assume that all of the light originates 
from the mass donor.  
Using the period--mass relations of 
Frank, King \& Raine (1992) and 
Warner (1995) together with the 4.46~h period of the system (MPW)
results in mass estimates for the donor star 
of 0.49~M$_{\sun}$ and 0.42~M$_{\sun}$, respectively. If the stars are 
normal main sequence stars, these masses imply a spectral type of K9-M1
(Allen 1973).  
Our upper limits in $I$  
require a   
M0V star with $A_V=1$ to be at a distance d$>5$~kpc and at d$>4$~kpc for 
$A_V=2$ (using the extinction relationship of Cardelli, Clayton \& Mathis 1989). 
Alternatively, if we make no assumption about the spectral 
type of the donor star, at a distance of 9~kpc with $A_V=2$ the spectral 
type of the
secondary would have to be
later than K3 in order not to be detected in $I$.

 \begin{table}
 \dummytable\label{t1}
 \end{table}

\subsection{X1658$-$298}

X1658$-$298 is a transient X-ray burst source discovered in 1976 by 
Lewin, Hoffmann \& Doty (1976).
Observations during a temporary brightening of the source in 1978 showed dips in the 
X-ray lightcurve. Detailed analysis of the combined 1976--1978 data set by Cominsky \&
Wood (1984, 1989) revealed that some of the dips are in fact eclipses of the 
central X-ray source by the mass donating star with a recurrence period of 7.1~hours.  
The dipping activity lasts for about 25\% of the 
orbital cycle followed by an eclipse of $\sim 15$~min duration. X1658$-$298 
entered an X-ray off--state in 1979 and has not
been detected in X-rays since.  

The optical counterpart was identified during the 1978 X-ray outburst 
with a faint ($V=18.3$), blue star (V2134~Oph) 
by Doxsey et al.\ (1979). Spectroscopic observations showed a blue continuum with 
emission lines of \ion{He}{2} $\lambda$4686 and the \ion{C}{3}/\ion{N}{3} 
$\lambda$4640/4650 blend (Canizares, McClintock \& Grindlay 1979).
In 1979 June, the counterpart was detected with $V=21.5$; about a month later it was 
undetectable with a magnitude limit of $V>23$ (Cominsky, Ossmann \& Lewin 1983). 
There is evidence that the source has been brightening gradually since then. 
Cowley, Hutchings \& Crampton (1988) mention obtaining a spectrum of the ``very faint
source'' in 1986.  
Shahbaz et al.\ (1996) display a featureless spectrum of V2134~Oph taken in 1988 
and estimate $V=20.7$ for the brightness of the counterpart.  
Navarro (1996) reported that spectra taken in 
1992/1993 showed H$\alpha$ in emission, but that simultaneous {\it VLA} and 
{\it Ginga} observations failed to detect any radio or X-ray flux. 

\begin{figure}[ht]
\epsscale{0.45}
\plotone{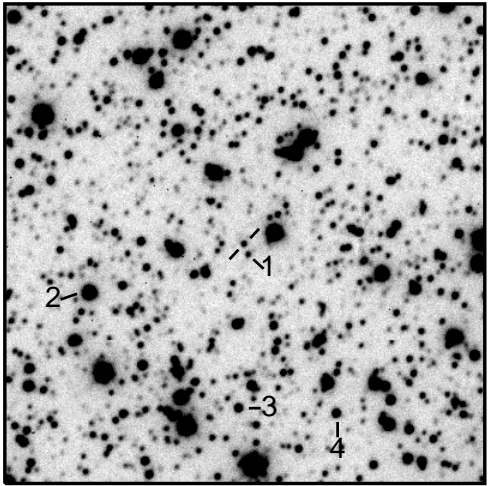}
\caption[]{600~s $I$ band exposure of the X1658$-$298 field.
The field size is 96\arcsec$\times$96\arcsec. North is up and east is to the left.
V2134~Oph, the counterpart of the X-ray source, is flagged. 
 \label{f-fc2}}
\end{figure}

We observed V2134~Oph on several occasions in April/May 1997 and found the 
source at a mean brightness of $V=19.54$, only $\sim$~1~mag fainter than when 
it is X-ray active. Since the only finding chart of X1658$-$298 was taken 
on photographic plates and is of limited quality, one of our $I$ frames
is displayed in Figure~\ref{f-fc2}. 
We performed
astrometry on our CCD frame to ensure that the observed star is indeed 
V2134~Oph and not a close companion unrelated to the X-ray source. Our 
position for V2134~Oph is 17:02:06.42, $-$29:56:44.33 (J2000) with an 
internal uncertainty
of 0.1\arcsec\ in each coordinate. Doxsey et al.\ (1979) list 
17:02:06.37, $-$29:56:43.23
for the counterpart (no error estimate is given), which only differs by 
1\arcsec\ in declination.  It therefore appears that the correct counterpart 
has been observed.
A comparison between our $B-V$ color 
($B-V=0.80$) and that of Doxsey et al.\ (1979, $B-V=0.37$) 
shows that the source is presently substantially redder than during outburst.

Cominsky \& Wood (1984) discuss the properties of possible secondary 
stars consistent with producing a 15~min eclipse in a 7.1~hour binary. 
The limiting cases are a 0.3~M$_{\sun}$ (M5) star that does not fill its Roche lobe
viewed at an inclination of $i=90^\circ$ and a 0.9~M$_{\sun}$ (G5) Roche lobe 
filling star
viewed at an inclination of $i=71.5^\circ$.  
Using the period--mass relations of 
Warner (1995) and Frank, King \& Raine (1992) 
results in mass estimates of 0.75--0.78~M$_{\sun}$ for the donor star, which
corresponds to a K0 main sequence star. Cominsky (1981) derived a distance 
of 15~kpc for X1658$-$298. At that distance, the small reddening ($E_{B-V}=0.3$) 
and the limit 
of $V>23$ imply that the secondary's spectral type is later than K2.     
A K star companion is very typical for a large amplitude X-ray transient.
If we assume a K3 star mass donor at 15~kpc it would contribute less than 5\%
to the current system brightness, so that most of the light is due to an 
accretion disk. This is supported by our dereddened colors, which are marginally
consistent (within the uncertainties) only with an F5 spectral type.  
Such a spectral type is
too luminous for the previously observed faint source states. 

\begin{figure}[ht]
\epsscale{0.80}
\plotone{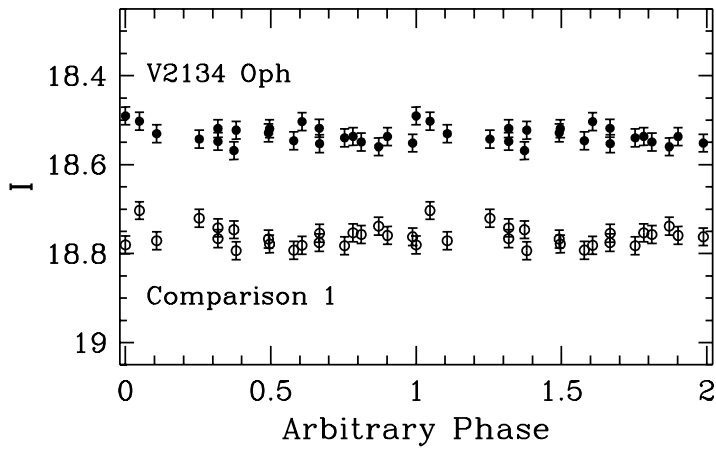}
\caption[]{$I$ band lightcurve of V2134~Oph, the 
optical counterpart to 
X1658$-$298, and comparison star 1 (see Figure~\ref{f-fc2}) folded on the 7.1~hour
period of X1658$-$298. 
The 1$\sigma$ error bars have been derived from the scatter
in the comparison lightcurves.  
\label{f-lc}}
\end{figure}

Our optical lightcurve, folded on the 7.1~hour 
orbital period, does not show any modulation across the binary cycle
(Figure~\ref{f-lc}). This is somewhat surprising since many transients in 
quiescence display photometric variability due to ellipsoidal variations.
However, these modulations could be 
masked by the presence of the apparently luminous accretion disk in 
X1658$-$298. An accretion disk with non--uniform illumination (e.g. with 
a hot spot) would also cause optical variability when viewed across the 
binary orbit. We place an upper limit of 0.02~mag (99\% confidence)
on the semi--amplitude of any 
sinusoidal modulation in the folded lightcurve. 
Unfortunately we cannot compare the outburst and quiescent variability 
characteristics, since there are no 
data available on the presence or absence of optical modulation during the 
1978 outburst. We do not 
observe any eclipses, but due to our limited phase coverage  
and the 
short duration of the eclipses they could have been easily missed. 

The optical behavior of X1658$-$298 is very different from other X-ray transients,
which are characterized by a steep increase in optical brightness during 
an X-ray outburst and roughly constant magnitudes in quiescence.  
X1658$-$298, in contrast, seems to be brightening gradually in the optical with no 
accompanying evidence for an increase in the X-ray emission.  
The RXTE ASM confirms that the source remains at $< 1$ mCrab 
during our observations. 
It is possible that the absence of X-ray emission, despite the indication 
for an accretion disk and activity in the system, is related to the apparent
high inclination of X1658$-$298. Changes in the accretion rate could have 
altered the structure in the outer parts of the accretion disk in such a way that  
it now permanently  
obscures the 
X-rays emanating from the central source. This would also block our view 
of the inner, hotter regions of the disk which is consistent with the 
redder colors of our observations.

\acknowledgments

We thank Bruce Margon 
for reading a draft of this paper and 
providing helpful comments.  This research  
has made use of the Simbad 
database, operated at CDS, Strasbourg, France.

\end{document}